\documentclass[12pt]{article}
\usepackage{amsmath}
\usepackage{graphicx}
\usepackage{amsfonts}
\usepackage{amssymb}
\begin{document}

\begin{center}
{\Large \textbf{Spherical gauge fields}}\footnote{Support provided by the
National Science Foundation under Grant 5-22968 and the Deutsche
Forschungsgemeinschaft}\\[1cm]

\textrm{Bu\={g}ra Borasoy}\footnote{email: borasoy@het.phast.umass.edu}%
\textrm{ and Dean Lee}\footnote{email: dlee@het.phast.umass.edu}

{\Large {\normalsize \textrm{University of Massachusetts}}}

{\Large {\normalsize \textrm{Amherst, MA 01003}}}

{\Large {\normalsize \textrm{\newline \vspace{24pt} {\small \parbox
{360pt}{We introduce the
spherical field formalism for free gauge fields.
We discuss the structure
of the spherical Hamiltonian for both general covariant
gauge and radial gauge and point out several new
features not present in the scalar field case. We then use
the evolution equations to compute
gauge-field and field-strength correlators.
[PACS numbers: 11.10.Kk, 11.15.Tk, 11.15.-q]} \vspace{10pt} }}}}
\end{center}

\section{Overview}

Spherical field theory is a new non-perturbative method for studying quantum
field theory. It was introduced in \cite{intro} to describe scalar boson
interactions. The method utilizes the spherical partial wave expansion to
reduce a general $d$-dimensional Euclidean functional integral into a set of
coupled one-dimensional, radial systems. The coupled one-dimensional systems
are then converted to differential equations which then can be solved using
standard numerical methods.\footnote{For free field theory these equations can
be solved exactly, as we will demonstrate here for gauge fields.} \ The
extension of the spherical field method to fermionic systems was described in
\cite{ferm}. In that analysis it was shown that the formalism avoids several
difficulties which appear in the lattice treatment of fermions. These include
fermion doubling, missing axial anomalies, and computational difficulties
arising from internal fermion loops. This finding suggests that the spherical
formalism could provide a useful method for studying gauge theories,
especially those involving fermions. As a small but important initial step in
this direction, we contribute the present work in which we introduce and
discuss the spherical field method for free gauge fields.

The basic formalism for spherical boson fields was described in \cite{intro}.
In this paper we will build on those results with most of our attention
devoted to new features resulting from the intrinsic spin of the gauge field.
We discuss the operator structure of the spherical Hamiltonian in detail,
using two-dimensional Euclidean gauge fields as an explicit example. Like
standard field theory gauge-fixing is essential in spherical field theory, and
we have chosen to consider general covariant gauge and radial
gauge\footnote{See \cite{leu} and references therein for a discussion of
radial gauge.}. In each case we derive the spherical Hamiltonian and use the
corresponding evolution equations to calculate the two-point correlators for
the gauge field and the gauge-invariant field strength. Free gauge fields in
higher dimensions can be described by a straightforward generalization of the
methods presented here. The application of spherical field theory to
non-perturbative interacting gauge systems and related issues are the subject
of current research.

\section{Covariant gauge}

In this section we derive the spherical field Hamiltonian for general
covariant gauge. We will use both polar and cartesian coordinates with the
following conventions:
\begin{equation}
\vec{t}=(t\cos\theta,t\sin\theta)=(t^{1},t^{2})=(x,y).
\end{equation}
In general covariant gauge the Euclidean functional integral is given by
\begin{equation}
\int\left(
%TCIMACRO{\tprod _{i}}%
%BeginExpansion
{\textstyle\prod_{i}}
%EndExpansion
\mathcal{D}A^{i}\right)  \exp\left[  \int_{0}^{\infty}dt\,L\right]
\end{equation}
where
\begin{equation}
L=\int d\theta\,t\left[  -\tfrac{1}{2}F^{12}F^{12}-\tfrac{1}{2\alpha}%
(\partial_{i}A^{i})^{2}\right]  . \label{lagr}%
\end{equation}
We can write the field strength $F^{12}$ as
\begin{align}
F^{12}  &  =\tfrac{1}{2i}\left[  \left(  \tfrac{\partial}{\partial x}%
+i\tfrac{\partial}{\partial y}\right)  (A^{x}-iA^{y})-\left(  \tfrac{\partial
}{\partial x}-i\tfrac{\partial}{\partial y}\right)  (A^{x}+iA^{y})\right] \\
&  =\tfrac{1}{\sqrt{2}i}\left[  e^{i\theta}\left(  \tfrac{\partial}{\partial
t}+\tfrac{i}{t}\tfrac{\partial}{\partial\theta}\right)  A^{+1}-e^{-i\theta
}\left(  \tfrac{\partial}{\partial t}-\tfrac{i}{t}\tfrac{\partial}%
{\partial\theta}\right)  A^{-1}\right] \nonumber
\end{align}
where
\begin{equation}
A^{x}\mp iA^{y}=\sqrt{2}A^{\pm1}.
\end{equation}
We now decompose $A^{\pm1}$ into partial waves\footnote{In our notation
$A_{n}^{\pm1}$ carries total spin quantum number $n\pm1$.}
\begin{equation}
A^{\pm1}=\tfrac{1}{\sqrt{2\pi}}\sum_{n=0,\pm1,\cdots}A_{n}^{\pm1}e^{in\theta}.
\end{equation}
Returning to our expression for the field strength, we have
\begin{equation}
F^{12}=\tfrac{1}{\sqrt{2\pi}}\tfrac{1}{\sqrt{2}i}\sum_{n=0,\pm1,\cdots
}e^{in\theta}\left(  F_{n}^{+1}-F_{n}^{-1}\right)  ,
\end{equation}
where
\begin{align}
F_{n}^{+1}  &  =\tfrac{\partial A_{n-1}^{+1}}{\partial t}-\tfrac{n-1}%
{t}A_{n-1}^{+1}\\
F_{n}^{-1}  &  =\tfrac{\partial A_{n+1}^{-1}}{\partial t}+\tfrac{n+1}%
{t}A_{n+1}^{-1}.
\end{align}
We can also express the gauge-fixing term in terms of $F_{n}^{\pm1}$,
\begin{equation}
\partial_{i}A^{i}=\tfrac{1}{\sqrt{2\pi}}\tfrac{1}{\sqrt{2}}\sum_{n=0,\pm
1,\cdots}e^{in\theta}\left(  F_{n}^{+1}+F_{n}^{-1}\right)  .
\end{equation}
With these changes the Lagrangian, $L$, is
\begin{equation}
\tfrac{t}{4}\sum_{n=0,\pm1,\cdots}\left(  F_{-n}^{+1}-F_{-n}^{-1}\right)
\left(  F_{n}^{+1}-F_{n}^{-1}\right)  -\tfrac{t}{4\alpha}\sum_{n=0,\pm
1,\cdots}\left(  F_{-n}^{+1}+F_{-n}^{-1}\right)  \left(  F_{n}^{+1}+F_{n}%
^{-1}\right)  .
\end{equation}
In \cite{intro} the spherical Hamiltonian for the scalar field was found by
direct application of the Feynman-Kac formula. This is also possible here, but
in view of the number of mixed terms (a result of the intrinsic spin degrees
of freedom) we find it easier to use the method of canonical quantization. Let
us define the conjugate momenta to the gauge fields,
\begin{align}
\pi_{n-1}^{+1}  &  =\frac{\delta L}{\delta\tfrac{\partial A_{n-1}^{+1}%
}{\partial t}}=\tfrac{t}{2}\left[  (1-\tfrac{1}{\alpha})F_{-n}^{+1}%
+(-1-\tfrac{1}{\alpha})F_{-n}^{-1}\right] \\
\pi_{n+1}^{-1}  &  =\frac{\delta L}{\delta\tfrac{\partial A_{n+1}^{-1}%
}{\partial t}}=\tfrac{t}{2}\left[  (-1-\tfrac{1}{\alpha})F_{-n}^{+1}%
+(1-\tfrac{1}{\alpha})F_{-n}^{-1}\right]  .
\end{align}
Following through with the canonical quantization procedure, we find a
Hamiltonian of the form%
\begin{equation}
H=\sum_{n=0,\pm1,\cdots}H_{n},
\end{equation}
where
\begin{align}
H_{n}  &  =-\tfrac{1}{4t}\left(  \pi_{-n+1}^{-1}-\pi_{-n-1}^{+1}\right)
\left(  \pi_{n-1}^{+1}-\pi_{n+1}^{-1}\right) \label{ham}\\
&  -\tfrac{\alpha}{4t}\left(  \pi_{-n+1}^{-1}+\pi_{-n-1}^{+1}\right)  \left(
\pi_{n-1}^{+1}+\pi_{n+1}^{-1}\right)  +\tfrac{n-1}{t}A_{n-1}^{+1}\pi
_{n-1}^{+1}-\tfrac{n+1}{t}A_{n+1}^{-1}\pi_{n+1}^{-1}.\nonumber
\end{align}
We obtain the corresponding Schr\"{o}dinger time evolution generator by making
the replacements
\begin{equation}
A_{n}^{\pm1}\rightarrow z_{n}^{\pm1},\qquad\pi_{n}^{\pm1}\rightarrow
\tfrac{\partial}{\partial z_{n}^{\pm1}}.
\end{equation}
We then find
\begin{align}
H_{n}  &  =-\tfrac{1}{4t}\left(  \tfrac{\partial}{\partial z_{-n+1}^{-1}%
}-\tfrac{\partial}{\partial z_{-n-1}^{+1}}\right)  \left(  \tfrac{\partial
}{\partial z_{n-1}^{+1}}-\tfrac{\partial}{\partial z_{n+1}^{-1}}\right) \\
&  -\tfrac{\alpha}{4t}\left(  \tfrac{\partial}{\partial z_{-n+1}^{-1}}%
+\tfrac{\partial}{\partial z_{-n-1}^{+1}}\right)  \left(  \tfrac{\partial
}{\partial z_{n-1}^{+1}}+\tfrac{\partial}{\partial z_{n+1}^{-1}}\right)
+\tfrac{n-1}{t}z_{n-1}^{+1}\tfrac{\partial}{\partial z_{n-1}^{+1}}-\tfrac
{n+1}{t}z_{n+1}^{-1}\tfrac{\partial}{\partial z_{n+1}^{-1}}.\nonumber
\end{align}
For the sake of future numerical calculations, it is convenient to re-express
$H$ in terms of real variables. Let us define
\begin{align}
u_{n}  &  =\tfrac{1}{2}\left(  z_{n+1}^{-1}+z_{-n-1}^{+1}+z_{n-1}%
^{+1}+z_{-n+1}^{-1}\right) \\
v_{n}  &  =\tfrac{1}{2}\left(  z_{n+1}^{-1}+z_{-n-1}^{+1}-z_{n-1}%
^{+1}-z_{-n+1}^{-1}\right) \\
x_{n}  &  =\tfrac{1}{2i}\left(  z_{n+1}^{-1}-z_{-n-1}^{+1}-z_{n-1}%
^{+1}+z_{-n+1}^{-1}\right) \\
y_{n}  &  =\tfrac{1}{2i}\left(  z_{n+1}^{-1}-z_{-n-1}^{+1}+z_{n-1}%
^{+1}-z_{-n+1}^{-1}\right)
\end{align}
for $n>0$ and%
\begin{align}
u_{0}  &  =\tfrac{1}{\sqrt{2}}\left(  z_{1}^{-1}+z_{-1}^{+1}\right) \\
x_{0}  &  =\tfrac{1}{\sqrt{2}i}\left(  z_{1}^{-1}-z_{-1}^{+1}\right)  .
\end{align}
The variables $u_{n}$ and $y_{n}$ correspond with combinations of$\ $radially
polarized gauge fields with total spin $\pm n,$ while $v_{n}$ and $x_{n}$
correspond with tangentially polarized gauge fields with total spin $\pm n$.
In terms of these new variables,
\begin{equation}
H=H_{0}+\sum_{n=1,2,\cdots}H_{n}^{\prime},
\end{equation}
where
\begin{equation}
H_{0}=-\tfrac{1}{2t}\tfrac{\partial^{2}}{\partial x_{0}^{2}}-\tfrac{\alpha
}{2t}\tfrac{\partial^{2}}{\partial u_{0}^{2}}-\tfrac{1}{t}\left(  u_{0}%
\tfrac{\partial}{\partial u_{0}}+x_{0}\tfrac{\partial}{\partial x_{0}}\right)
, \label{ze}%
\end{equation}
and%
\begin{align}
H_{n}^{\prime}=  &  -\tfrac{1}{2t}\left(  \tfrac{\partial^{2}}{\partial
v_{n}^{2}}+\tfrac{\partial^{2}}{\partial x_{n}^{2}}\right)  -\tfrac{\alpha
}{2t}\left(  \tfrac{\partial^{2}}{\partial u_{n}^{2}}+\tfrac{\partial^{2}%
}{\partial y_{n}^{2}}\right) \\
&  -\tfrac{n}{t}\left(  u_{n}\tfrac{\partial}{\partial v_{n}}+v_{n}%
\tfrac{\partial}{\partial u_{n}}+x_{n}\tfrac{\partial}{\partial y_{n}}%
+y_{n}\tfrac{\partial}{\partial x_{n}}\right) \nonumber\\
&  -\tfrac{1}{t}\left(  u_{n}\tfrac{\partial}{\partial u_{n}}+v_{n}%
\tfrac{\partial}{\partial v_{n}}+x_{n}\tfrac{\partial}{\partial x_{n}}%
+y_{n}\tfrac{\partial}{\partial y_{n}}\right)  .\nonumber
\end{align}
Despite the somewhat complicated form, we know that the spectrum of $H$ should
resemble that of harmonic oscillators. To see this explicitly let us define
the following ladder operators:
\begin{equation}
U_{n}^{+}=-\tfrac{1}{\sqrt{2}}\left(  \tfrac{\partial}{\partial u_{n}}%
+\tfrac{\partial}{\partial v_{n}}\right)  ,\quad U_{n}^{-}=\tfrac{1}{\sqrt{2}%
}\left(  \tfrac{2\alpha-n+\alpha n}{4(n+1)}\tfrac{\partial}{\partial u_{n}%
}+\tfrac{2+n-\alpha n}{4(n+1)}\tfrac{\partial}{\partial v_{n}}+u_{n}%
+v_{n}\right)  ,
\end{equation}%
\begin{equation}
V_{n}^{+}=\tfrac{1}{\sqrt{2}}\left(  \tfrac{-2\alpha-n+\alpha n}{4(n-1)}%
\tfrac{\partial}{\partial u_{n}}+\tfrac{2-n+\alpha n}{4(n-1)}\tfrac{\partial
}{\partial v_{n}}+u_{n}-v_{n}\right)  ,\quad V_{n}^{-}=\tfrac{1}{\sqrt{2}%
}\left(  \tfrac{\partial}{\partial u_{n}}-\tfrac{\partial}{\partial v_{n}%
}\right)  ,
\end{equation}%
\begin{equation}
X_{n}^{+}=-\tfrac{1}{\sqrt{2}}\left(  \tfrac{\partial}{\partial y_{n}}%
+\tfrac{\partial}{\partial x_{n}}\right)  ,\quad X_{n}^{-}=\tfrac{1}{\sqrt{2}%
}\left(  \tfrac{2\alpha-n+\alpha n}{4(n+1)}\tfrac{\partial}{\partial y_{n}%
}+\tfrac{2+n-\alpha n}{4(n+1)}\tfrac{\partial}{\partial x_{n}}+y_{n}%
+x_{n}\right)  ,
\end{equation}%
\begin{equation}
Y_{n}^{+}=\tfrac{1}{\sqrt{2}}\left(  \tfrac{-2\alpha-n+\alpha n}{4(n-1)}%
\tfrac{\partial}{\partial y_{n}}+\tfrac{2-n+\alpha n}{4(n-1)}\tfrac{\partial
}{\partial x_{n}}+y_{n}-x_{n}\right)  ,\quad Y_{n}^{-}=\tfrac{1}{\sqrt{2}%
}\left(  \tfrac{\partial}{\partial y_{n}}-\tfrac{\partial}{\partial x_{n}%
}\right)  ,
\end{equation}
for $n>1$ and%
\begin{align}
U_{0}^{+}  &  =-\tfrac{1}{\sqrt{2}}\tfrac{\partial}{\partial u_{0}},\quad
U_{0}^{-}=\tfrac{\alpha}{\sqrt{2}}\tfrac{\partial}{\partial u_{0}}+\sqrt
{2}u_{0},\\
X_{0}^{+}  &  =-\tfrac{1}{\sqrt{2}}\tfrac{\partial}{\partial x_{0}},\quad
X_{0}^{-}=\tfrac{1}{\sqrt{2}}\tfrac{\partial}{\partial x_{0}}+\sqrt{2}x_{0}.
\end{align}
We have constructed these operators so that
\begin{equation}
\left[  U_{n}^{-},U_{n}^{+}\right]  =\left[  V_{n}^{-},V_{n}^{+}\right]
=\left[  X_{n}^{-},X_{n}^{+}\right]  =\left[  Y_{n}^{-},Y_{n}^{+}\right]  =1
\end{equation}
for $n>1$ and%
\begin{equation}
\left[  U_{0}^{-},U_{0}^{+}\right]  =\left[  X_{0}^{-},X_{0}^{+}\right]  =1.
\end{equation}
All other commutators involving these operators vanish. We can now rewrite
$H_{0}$ and $H_{n}^{\prime}$ (for $n>1$) as
\begin{equation}
H_{0}=\tfrac{1}{t}U_{0}^{+}U_{0}^{-}+\tfrac{1}{t}X_{0}^{+}X_{0}^{-}+\tfrac
{2}{t},
\end{equation}%
\begin{equation}
H_{n}^{\prime}=\tfrac{n+1}{t}U_{n}^{+}U_{n}^{-}+\tfrac{n-1}{t}V_{n}^{+}%
V_{n}^{-}+\tfrac{n+1}{t}X_{n}^{+}X_{n}^{-}+\tfrac{n-1}{t}Y_{n}^{+}Y_{n}%
^{-}+\tfrac{2n+2}{t}.
\end{equation}
\qquad We now see that $H_{0}$ is equivalent to two harmonic oscillators with
level-spacing $\tfrac{1}{t},$ and $H_{n}^{\prime}$ is equivalent to four
harmonic oscillators, two with spacing $\tfrac{n+1}{t}$ and two with spacing
$\tfrac{n-1}{t}$. It is important to note that $H_{1}^{\prime}$, however, is
quite different. The $s$-wave configurations are independent of $\theta$, and
so $H_{1}^{\prime}$ does not have terms of the form $z_{0}^{+1}\tfrac
{\partial}{\partial z_{0}^{+1}}$ and $z_{0}^{-1}\tfrac{\partial}{\partial
z_{0}^{-1}}$. Furthermore the gauge fields are massless and so $H_{1}^{\prime
}$ depends only on the derivatives of $z_{0}^{+1}$ and $z_{0}^{-1}$. One
consequence of this is that the spectrum of $H_{1}^{\prime}$ is continuous,
and calculations involving $H_{1}^{\prime}$ and relevant boundary conditions
must be done carefully. We will discuss these issues later in our analysis of
correlation functions.

In preparation for our calculation of gauge-field correlators, let us now
couple an external source $\mathcal{\vec{J}}$ to the gauge field,%
\begin{equation}
L=\int d\theta\,t\left[  -\tfrac{1}{2}F^{12}F^{12}-\tfrac{1}{2\alpha}%
(\partial_{i}A^{i})^{2}+\vec{A}\cdot\mathcal{\vec{J}}\right]  .
\end{equation}
The new spherical field Hamiltonian is then%
\begin{equation}
H(\mathcal{\vec{J}})=H-\sum_{n=0,\pm1,\cdots}\left(  z_{n}^{+1}\mathcal{J}%
_{-n}^{-1}+z_{n}^{-1}\mathcal{J}_{-n}^{+1}\right)  .
\end{equation}
As noted in \cite{intro} the vacuum persistence amplitude is given by%
\begin{equation}
Z(\mathcal{\vec{J}})\varpropto\lim_{\substack{t_{\min}\longrightarrow
0^{+}\\t_{\max}\longrightarrow\infty}}\left\langle b\right|  T\exp\left[
-\int_{t_{\min}}^{t_{\max}}dt\,H(\mathcal{\vec{J}})\right]  \left|
a\right\rangle , \label{gf}%
\end{equation}
where $\left|  a\right\rangle $ and $\left|  b\right\rangle $ are any states
satisfying certain criteria. These criteria are that $\left|  a\right\rangle $
is constant with respect to the s-wave variables $z_{0}^{-1}$, $z_{0}^{+1}$
and has non-zero overlap with the ground state of $H$ as $t\rightarrow
0^{\text{+}}$, and $\left|  b\right\rangle $ has non-zero overlap with the
ground state of $H$ as $t\rightarrow\infty$.\footnote{One caveat here is that
$\left|  a\right\rangle $ must lie in a function space over which the spectrum
of $H$ is bounded below.} Because the spectrum of $H_{1}^{\prime}$ is
continuous, in numerical computations it is useful to include a small
regulating mass, $\mu$, for the gauge fields and then take the limit
$\mu\rightarrow0$.

\section{Radial gauge}

We now derive the spherical field Hamiltonian for radial gauge. We take the
gauge-fixing reference point, $\vec{t}_{0}$, to be the origin,%
\begin{equation}
(\vec{t}-\vec{t}_{0})\cdot\vec{A}=\vec{t}\cdot\vec{A}=0.
\end{equation}
We expect this gauge-fixing scheme to be convenient in spherical field theory
calculations for several reasons. One is that non-abelian ghost fields in
radial gauge decouple, as they do in axial gauge. In contrast with axial
gauge, however, radial gauge also preserves rotational symmetry. As we will
see, the spherical Hamiltonian and correlation functions in radial gauge are
relatively simple. Since%

\begin{equation}
\vec{t}\cdot\vec{A}=\tfrac{t}{\sqrt{2}}\left[  e^{i\theta}A^{+1}+e^{-i\theta
}A^{-1}\right]  ,
\end{equation}
we can impose the gauge-fixing condition by setting
\begin{equation}
A^{-1}=-e^{2i\theta}A^{+1}.
\end{equation}
With this constraint we express the field strength as
\begin{equation}
F^{12}=\tfrac{1}{i\sqrt{\pi}}\sum_{n=0,\pm1,\cdots}e^{in\theta}\left[
\tfrac{\partial A_{n-1}^{+1}}{\partial t}+\tfrac{1}{t}A_{n-1}^{+1}\right]  .
\end{equation}
The radial-gauge Lagrangian is then
\begin{align}
L  &  =\int d\theta\,t\left[  -\tfrac{1}{2}F^{12}F^{12}\right] \\
&  =t\sum_{n=0,\pm1,\cdots}\left[  \tfrac{\partial A_{-n-1}^{+1}}{\partial
t}+\tfrac{1}{t}A_{-n-1}^{+1}\right]  \left[  \tfrac{\partial A_{n-1}^{+1}%
}{\partial t}+\tfrac{1}{t}A_{n-1}^{+1}\right]  .\nonumber
\end{align}
We again follow the canonical quantization procedure. The conjugate momenta to
the gauge fields are
\begin{equation}
\pi_{n-1}^{+1}=\frac{\delta L}{\delta\frac{\partial A_{n-1}^{+1}}{\partial t}%
}=2t\left[  \tfrac{\partial A_{-n-1}^{+1}}{\partial t}+\tfrac{1}{t}%
A_{-n-1}^{+1}\right]  ,
\end{equation}
and the radial-gauge Hamiltonian has the form
\begin{equation}
H=\sum_{n=0,\pm1,\cdots}\left[  \tfrac{1}{4t}\pi_{n-1}^{+1}\pi_{-n-1}%
^{+1}-\tfrac{1}{t}A_{n-1}^{+1}\pi_{n-1}^{+1}\right]  .
\end{equation}
In the Schr\"{o}dinger language the Hamiltonian becomes%

\begin{equation}
H=\sum_{n=0,\pm1,\cdots}\left[  \tfrac{1}{4t}\tfrac{\partial}{\partial
z_{n-1}^{+1}}\tfrac{\partial}{\partial z_{-n-1}^{+1}}-\tfrac{1}{t}z_{n-1}%
^{+1}\tfrac{\partial}{\partial z_{n-1}^{+1}}\right]  .
\end{equation}
As before we now define real variables,
\begin{align}
x_{n}  &  =\tfrac{i}{\sqrt{2}}\left(  z_{n-1}^{+1}+z_{-n-1}^{+1}\right) \\
v_{n}  &  =\tfrac{1}{\sqrt{2}}\left(  z_{n-1}^{+1}-z_{-n-1}^{+1}\right)
\end{align}
for $n>0$ and%
\begin{equation}
x_{0}=i\cdot z_{-1}^{+1}.
\end{equation}
Our Hamiltonian can be re-expressed as
\begin{equation}
H=H_{0}+\sum_{n=1,2,\cdots}H_{n}^{\prime},
\end{equation}
where
\begin{equation}
H_{0}=-\tfrac{1}{4t}\tfrac{\partial^{2}}{\partial x_{0}^{2}}-\tfrac{1}{t}%
x_{0}\tfrac{\partial}{\partial x_{0}},
\end{equation}%
\begin{equation}
H_{n}^{^{\prime}}=-\tfrac{1}{4t}\left[  \tfrac{\partial^{2}}{\partial
x_{n}^{2}}+\tfrac{\partial^{2}}{\partial v_{n}^{2}}\right]  -\tfrac{1}%
{t}\left[  x_{n}\tfrac{\partial}{\partial x_{n}}+v_{n}\tfrac{\partial
}{\partial v_{n}}\right]  .
\end{equation}
Let us now define ladder operators
\begin{equation}
X_{n}^{+}=\tfrac{1}{2}\tfrac{\partial}{\partial x_{n}},\quad X_{n}^{-}%
=-\tfrac{1}{2}\tfrac{\partial}{\partial x_{n}}-2x_{n},
\end{equation}
for $n\geq0$ and
\begin{equation}
V_{n}^{+}=\tfrac{1}{2}\tfrac{\partial}{\partial v_{n}},\quad V_{n}^{-}%
=-\tfrac{1}{2}\tfrac{\partial}{\partial v_{n}}-2v_{n},
\end{equation}
for $n>0.$ These ladder operators satisfy the relations
\begin{equation}
\left[  X_{n}^{-},X_{n}^{+}\right]  =\left[  V_{n}^{-},V_{n}^{+}\right]  =1,
\end{equation}
while all other commutators vanish. In terms of these operators we have
\begin{equation}
H_{0}=\tfrac{1}{t}X_{0}^{+}X_{0}^{-}+\tfrac{1}{t},
\end{equation}%
\begin{equation}
H_{n}^{\prime}=\tfrac{1}{t}X_{n}^{+}X_{n}^{-}+\tfrac{1}{t}V_{n}^{+}V_{n}%
^{-}+\tfrac{2}{t}.
\end{equation}
We note the radial gauge constraint has removed the continuous spectrum from
the $s$-wave sector, and the spectrum of $H$ is purely discrete. Furthermore
the splitting between energy levels of $H_{n}^{\prime}$ is independent of
$n$.\footnote{The reason is that in radial gauge the gauge fields are
tangentially polarized, and so purely tangential excitations in two dimensions
do not contribute to the field strength $F^{12}$.}

We now couple an external source $\mathcal{\vec{J}}$ to the gauge field,%
\begin{equation}
L=\int d\theta\,t\left[  -\tfrac{1}{2}F^{12}F^{12}+\vec{A}\cdot\mathcal{\vec
{J}}\right]  .
\end{equation}
The new Hamiltonian is then%
\begin{equation}
H(\mathcal{\vec{J}})=H-\sum_{n=0,\pm1,\cdots}z_{n}^{+1}\left(  \mathcal{J}%
_{-n}^{-1}+\mathcal{J}_{-n-2}^{+1}\right)  .
\end{equation}
The vacuum persistence amplitude in radial gauge is given by%
\begin{equation}
Z(\mathcal{\vec{J}})\varpropto\lim_{\substack{t_{\min}\longrightarrow
0^{+}\\t_{\max}\longrightarrow\infty}}\left\langle b\right|  T\exp\left[
-\int_{t_{\min}}^{t_{\max}}dt\,H(\mathcal{\vec{J}})\right]  \left|
a\right\rangle ,
\end{equation}
where $\left|  a\right\rangle $ has non-zero overlap with the ground state of
$H$ as $t\rightarrow0^{\text{+}}$, and $\left|  b\right\rangle $ has non-zero
overlap with the ground state of $H$ as $t\rightarrow\infty$. Since the
level-spacing of $H_{1}^{\prime}$ diverges as $t\rightarrow0^{\text{+}}$ only
the ground state projection of $H_{1}^{\prime}$ at $t=0$ contributes, and
$\left|  a\right\rangle $ no longer needs to be constant with respect\ to the
$s$-wave variables $z_{0}^{\pm1}$. This a rather important point since, as we
recall from \cite{intro}, the constant $s$-wave boundary condition at $t=0$
follows from the fact that the value of field at the origin is not
constrained. This is however not true in radial gauge as a result of the
gauge-fixing constraint.

\section{Gauge-field correlators}

In this section we calculate two-point gauge-field correlators using the
spherical field formalism. This calculation can be done in several ways,
including numerically. For future applications, however, it is useful to have
exact expressions for use in perturbative calculations (e.g., for evaluating
counterterms). Here we will obtain results by decomposing the fields as a
combination of the ladder operators. Let us start with radial gauge. We have%
\begin{align}
z_{n-1}^{+1}  &  =\tfrac{1}{2\sqrt{2}}\left(  iX_{n}^{+}+iX_{n}^{-}-V_{n}%
^{+}-V_{n}^{-}\right) \\
z_{-n-1}^{+1}  &  =\tfrac{1}{2\sqrt{2}}\left(  iX_{n}^{+}+iX_{n}^{-}+V_{n}%
^{+}+V_{n}^{-}\right)
\end{align}
for $n>0$ and%
\begin{equation}
z_{-1}^{+1}=\tfrac{i}{2}\left(  X_{0}^{+}+X_{0}^{-}\right)  .
\end{equation}
We would like to calculate the correlation function,%
\begin{equation}
f_{n-1,n^{\prime}-1}^{rad}(t,t^{\prime})=\left\langle 0\right|  A_{n-1}%
^{+1}(t)A_{n^{\prime}-1}^{+1}(t^{\prime})\left|  0\right\rangle _{rad}.
\end{equation}
By angular momentum conservation $f_{n-1,n^{\prime}-1}^{rad}$ vanishes unless
$n^{\prime}=-n$. Also
\begin{equation}
f_{n-1,n^{\prime}-1}^{rad}(t,t^{\prime})=f_{n^{\prime}-1,n-1}^{rad}(t^{\prime
},t),
\end{equation}
and so without loss of generality it suffices to consider $f_{n-1,-n-1}^{rad}$
for $n\geq0$. For typographical convenience let us define%
\begin{equation}
\left\langle \left\{  F\right\}  _{t}\left\{  G\right\}  _{t^{\prime}%
}\right\rangle =\,\theta(t-t^{\prime})\tfrac{\left\langle b\right|
U(\infty,t)FU(t,t^{\prime})GU(t^{\prime},0)\left|  a\right\rangle
}{\left\langle b\right|  U(\infty,0)\left|  a\right\rangle }+\theta(t^{\prime
}-t)\tfrac{\left\langle b\right|  U(\infty,t^{\prime})GU(t^{\prime
},t)FU(t,0)\left|  a\right\rangle }{\left\langle b\right|  U(\infty,0)\left|
a\right\rangle },
\end{equation}
where
\begin{equation}
U(t_{2},t_{1})=T\exp\left[  -\int_{t_{1}}^{t_{2}}dt\,H\right]  .
\end{equation}
For $n\neq0$, we have%
\begin{equation}
f_{n-1,-n-1}^{rad}=\tfrac{1}{8}\left\langle \left\{  iX_{n}^{+}+iX_{n}%
^{-}-V_{n}^{+}-V_{n}^{-}\right\}  _{t}\left\{  iX_{n}^{+}+iX_{n}^{-}+V_{n}%
^{+}+V_{n}^{-}\right\}  _{t^{\prime}}\right\rangle .
\end{equation}
Using the commutation properties of these ladder operators with $H(t)$, we
find%
\begin{equation}
f_{n-1,-n-1}^{rad}=-\tfrac{1}{4}\left[  \theta(t^{\prime}-t)\tfrac
{t}{t^{\prime}}+\theta(t-t^{\prime})\tfrac{t^{\prime}}{t}\right]  .
\end{equation}
For the special case $n=0,$ we obtain
\begin{equation}
f_{-1,-1}^{rad}=\tfrac{1}{4}\left\langle \left\{  iX_{0}^{+}+iX_{0}%
^{-}\right\}  _{t}\left\{  iX_{0}^{+}+iX_{0}^{-}\right\}  _{t^{\prime}%
}\right\rangle =-\tfrac{1}{4}\left[  \theta(t^{\prime}-t)\tfrac{t}{t^{\prime}%
}+\theta(t-t^{\prime})\tfrac{t^{\prime}}{t}\right]  .
\end{equation}
These correlation functions are in agreement with results we obtain by
decomposing the following expression \cite{leu} into partial waves:
\begin{align}
&  \left\langle 0\right|  A^{i}(\vec{x})A^{j}(\vec{y})\left|  0\right\rangle
_{rad}\label{rg}\\
&  =\tfrac{1}{4\pi}\lim_{\varepsilon\rightarrow0^{+}}\left[
\begin{array}
[c]{c}%
\delta^{ij}\log\tfrac{\left(  \vec{x}-\vec{y}\right)  ^{2}}{L^{2}}%
-\partial_{i}^{x}\int_{0}^{1}ds\,x^{j}\log\tfrac{\left(  s\vec{x}-\vec
{y}\right)  ^{2}+\varepsilon}{L^{2}}-\partial_{j}^{y}\int_{0}^{1}dt\,y^{i}%
\log\tfrac{\left(  \vec{x}-t\vec{y}\right)  ^{2}+\varepsilon}{L^{2}}\\
+\partial_{i}^{x}\partial_{j}^{y}\int_{0}^{1}ds\int_{0}^{1}dt\,\vec{x}%
\cdot\vec{y}\log\tfrac{\left(  s\vec{x}-t\vec{y}\right)  ^{2}+\varepsilon
}{L^{2}}%
\end{array}
\right]  .\nonumber
\end{align}
The length scale $L$ is used to render the argument of the logarithm
dimensionless. Its purpose, however, is only cosmetic since the gauge-field
correlator is not infrared divergent in radial gauge and the dependence on $L$
cancels.\footnote{The radial gauge constraint $A^{-1}=-e^{2i\theta}A^{+1}$
pairs $s$-wave configurations with $d$-wave configurations. Since the $d$-wave
is not infrared divergent, neither is the $s$-wave.}

These same methods can be applied to gauge-field correlators in covariant
gauge. We find
\begin{align}
\left\langle 0\right|  A_{n-1}^{+1}(t)A_{-n-1}^{+1}(t^{\prime})\left|
0\right\rangle _{cov}  &  =\left\langle 0\right|  A_{-n+1}^{-1}(t)A_{n+1}%
^{-1}(t^{\prime})\left|  0\right\rangle _{cov}\\
&  =\tfrac{\alpha-1}{4}\delta_{n,0}\left[  \theta(t-t^{\prime})\tfrac
{t^{\prime}}{t}+\theta(t^{\prime}-t)\tfrac{t}{t^{\prime}}\right]  \nonumber\\
&  -\tfrac{\alpha-1}{4}\left[  \theta(t-t^{\prime})\delta_{n,-1}+\theta(t^{\prime}-t)
\delta_{n,1}\right]  .\nonumber
\end{align}
and for $n\neq1$,
\begin{equation}
\left\langle 0\right|  A_{n-1}^{+1}(t)A_{-n+1}^{-1}(t^{\prime})\left|
0\right\rangle _{cov}=\tfrac{\alpha+1}{4\left|  n-1\right|  }\left[
\theta(t-t^{\prime})\left(  \tfrac{t^{\prime}}{t}\right)  ^{\left|
n-1\right|  }+\theta(t^{\prime}-t)\left(  \tfrac{t}{t^{\prime}}\right)
^{\left|  n-1\right|  }\right]  .
\end{equation}
In covariant gauge the $s$-wave correlator is infrared divergent,
\begin{equation}
\left\langle 0\right|  A_{0}^{+1}(t)A_{0}^{-1}(t^{\prime})\left|
0\right\rangle _{cov}=-\tfrac{\alpha+1}{2}\left[  \theta(t-t^{\prime}%
)\log\frac{t}{L}+\theta(t^{\prime}-t)\log\frac{t^{\prime}}{L}\right]  ,
\end{equation}
where $\log L$ is infinite.\ This divergence is specific to two-dimensional
gauge fields and does not occur in higher dimensions. If we include a
regulating gauge-field mass, $\mu$, we find that $L$ scales as $1/\mu$ as
$\mu\rightarrow0.$ These correlation functions are in agreement with the
results we obtain by decomposing the following known expression into partial
waves:
\begin{equation}
\left\langle 0\right|  A^{i}(\vec{x})A^{j}(\vec{y})\left|  0\right\rangle
_{cov}=-\tfrac{1}{4\pi}\left[  \tfrac{\alpha+1}{2}\delta^{ij}\log
\tfrac{\left(  \vec{x}-\vec{y}\right)  ^{2}}{L^{2}}+\tfrac{(\alpha
-1)(x-y)^{i}(x-y)^{j}}{\left(  \vec{x}-\vec{y}\right)  ^{2}}\right]  .
\end{equation}

\section{Gauge-invariant correlators}

Let us now consider the two-point correlator of the gauge-invariant field
strength $F^{12}$. We can calculate the $F^{12}$ correlator by differentiating
the gauge-field correlators calculated in the previous section, but it is
instructive to redo the calculation by coupling a source to $F^{12}$. This
time we describe the calculation in detail for covariant gauge. The same
calculation can be done for radial gauge using similar methods. Let us start
by quoting the result we expect. From free field theory we know%
\begin{equation}
\left\langle 0\right|  F^{12}(\vec{t})F^{12}(\vec{t}^{\prime})\left|
0\right\rangle =\delta^{2}(\vec{t}-\vec{t}^{\prime}).
\end{equation}
The $F^{12}$ correlator has a simple local structure, a consequence of the
fact that in two dimensions gauge fields can be decomposed into scalar and
longitudinal polarizations (borrowing Minkowski space terminology). There are
no transverse polarizations to produce non-local contributions to the $F^{12}$
correlator. We will see this happen explicitly in the calculations to follow.

The two-point correlator for partial waves of $F^{12}$ is given by
\begin{equation}
\left\langle 0\right|  F_{n}^{12}(t)F_{n^{\prime}}^{12}(t^{\prime})\left|
0\right\rangle =\tfrac{1}{2\pi}\int d\theta d\theta^{\prime}e^{-in\theta
}e^{-in^{\prime}\theta^{\prime}}\left\langle 0\right|  F^{12}(\vec{t}%
)F^{12}(\vec{t}^{\prime})\left|  0\right\rangle , \label{co}%
\end{equation}
and we deduce%
\begin{equation}
\left\langle 0\right|  F_{n}^{12}(t)F_{n^{\prime}}^{12}(t^{\prime})\left|
0\right\rangle =\tfrac{1}{t}\delta_{n,-n^{\prime}}\delta(t-t^{\prime}).
\label{corr}%
\end{equation}
Let us now reproduce this result using the spherical field method. We return
to the covariant gauge Lagrangian and couple a source $\mathcal{K}$ to
$F^{12}$,%

\begin{equation}
L=\int d\theta\,t\left[  -\tfrac{1}{2}F^{12}F^{12}-\tfrac{1}{2\alpha}%
(\partial_{i}A^{i})^{2}+F^{12}\mathcal{K}\right]  .
\end{equation}
The conjugate momenta are now
\begin{align}
\pi_{n-1}^{+1}  &  =\tfrac{t}{2}\left[  (1-\tfrac{1}{\alpha})F_{-n}%
^{+1}+(-1-\tfrac{1}{\alpha})F_{-n}^{-1}-i\sqrt{2}\mathcal{K}_{-n}\right] \\
\pi_{n+1}^{-1}  &  =\tfrac{t}{2}\left[  (-1-\tfrac{1}{\alpha})F_{-n}%
^{+1}+(1-\tfrac{1}{\alpha})F_{-n}^{-1}+i\sqrt{2}\mathcal{K}_{-n}\right]  ,
\end{align}
and the new Hamiltonian is%
\begin{equation}
H(\mathcal{K})=\sum_{n=0,\pm1,\cdots}H_{n}(\mathcal{K}),
\end{equation}
where
\begin{equation}
H_{n}(\mathcal{K})=H_{n}(0)+\tfrac{i}{\sqrt{2}}\mathcal{K}_{n}\pi_{n-1}%
^{+1}-\tfrac{i}{\sqrt{2}}\mathcal{K}_{n}\pi_{n+1}^{-1}-\tfrac{t}{2}%
\mathcal{K}_{-n}\mathcal{K}_{n}.
\end{equation}
In the Schr\"{o}dinger language we have
\begin{equation}
H_{n}(\mathcal{K})=H_{n}(0)+\tfrac{i}{\sqrt{2}}\mathcal{K}_{n}(-\tfrac
{n}{\left|  n\right|  }\tfrac{\partial}{\partial v_{\left|  n\right|  }%
}+i\tfrac{\partial}{\partial x_{\left|  n\right|  }})-\tfrac{t}{2}%
\mathcal{K}_{-n}\mathcal{K}_{n}%
\end{equation}
for $n\neq0$ and%
\begin{equation}
H_{0}(\mathcal{K})=H_{0}(0)-\mathcal{K}_{0}\tfrac{\partial}{\partial x_{0}%
}-\tfrac{t}{2}\mathcal{K}_{0}\mathcal{K}_{0}.
\end{equation}
The vacuum persistence amplitude is
\begin{equation}
Z(\mathcal{K})=\left\langle b\right|  T\exp\left[  -\int_{0}^{\infty
}dt\,H(\mathcal{K})\right]  \left|  a\right\rangle ,
\end{equation}
and we evaluate the $F_{n}^{12}$ correlator by functional differentiation with
respect to $\mathcal{K}$,
\begin{equation}
\left\langle 0\right|  F_{n}^{12}(t)F_{n^{\prime}}^{12}(t^{\prime})\left|
0\right\rangle =\left.  \tfrac{1}{Z(0)}\tfrac{\partial}{t\partial
\mathcal{K}_{-n}(t)}\tfrac{\partial}{t^{\prime}\partial\mathcal{K}%
_{-n^{\prime}}(t^{\prime})}Z(\mathcal{K})\right|  _{\mathcal{K}=0}.
\end{equation}
It is clear from angular momentum conservation that this correlator vanishes
unless $n=-n^{\prime}$, and so it suffices to compute%
\begin{equation}
\left\langle 0\right|  F_{n}^{12}(t)F_{-n}^{12}(t^{\prime})\left|
0\right\rangle . \label{v}%
\end{equation}
Since (\ref{v}) is symmetric under the interchange $n,t\leftrightarrow
-n,t^{\prime}$, we can also restrict $n\geq0$. Differentiating with respect to
the sources, we obtain, for $n>0$,
\begin{equation}
\left\langle 0\right|  F_{n}^{12}(t)F_{-n}^{12}(t^{\prime})\left|
0\right\rangle =-\tfrac{1}{2tt^{\prime}}\left\langle \left\{  \tfrac{\partial
}{\partial v_{n}}+i\tfrac{\partial}{\partial x_{n}}\right\}  _{t}\left\{
-\tfrac{\partial}{\partial v_{n}}+i\tfrac{\partial}{\partial x_{n}}\right\}
_{t^{\prime}}\right\rangle +\tfrac{1}{t}\delta(t-t^{\prime}). \label{b}%
\end{equation}
When $n>1$ we can write $\pm\tfrac{\partial}{\partial v_{n}}+i\tfrac{\partial
}{\partial x_{n}}$ as a linear combination of $U_{n}^{+}$, $V_{n}^{-}$,
$X_{n}^{+}$, and $Y_{n}^{-}$. These ladder operators are, however, acting in
four different spaces. The matrix element of the operator%
\begin{equation}
U(\infty,t)(\tfrac{\partial}{\partial v_{n}}+i\tfrac{\partial}{\partial x_{n}%
})U(t,t^{\prime})(-\tfrac{\partial}{\partial v_{n}}+i\tfrac{\partial}{\partial
x_{n}})U(t^{\prime},0)
\end{equation}
from the ground state at $t=0$ to the ground state at $t=\infty$
vanishes.\footnote{As noted before, this is due to the fact that in two
dimensions there are no transverse polarizations.} Consequently for $n>1$ only
the delta function contributes to the correlation function in (\ref{b}). The
same arguments apply for the case $n=0,$ and only the delta function
contributes here as well.

We now turn to the special case $n=1.$ The relevant part of the Hamiltonian
is
\begin{align}
H_{1}^{\prime}  &  =-\tfrac{1}{2t}\left(  \tfrac{\partial^{2}}{\partial
v_{1}^{2}}+\tfrac{\partial^{2}}{\partial x_{1}^{2}}\right)  -\tfrac{\alpha
}{2t}\left(  \tfrac{\partial^{2}}{\partial u_{1}^{2}}+\tfrac{\partial^{2}%
}{\partial y_{1}^{2}}\right) \\
&  -\tfrac{1}{t}\left[  \left(  u_{1}+v_{1}\right)  \left(  \tfrac{\partial
}{\partial u_{1}}+\tfrac{\partial}{\partial v_{1}}\right)  +\left(
x_{1}+y_{1}\right)  \left(  \tfrac{\partial}{\partial x_{1}}+\tfrac{\partial
}{\partial y_{1}}\right)  \right]  .\nonumber
\end{align}
The combinations $u_{1}-v_{1}$ and $x_{1}-y_{1}$ correspond with linear
combinations of the s-wave variables $z_{0}^{+1}$ and $z_{0}^{-1}.$ The
initial configuration $\left|  a\right\rangle $ at $t=0$ is constant with
respect to $z_{0}^{+1}$ and $z_{0}^{-1}$, and therefore constant with respect
to $u_{1}-v_{1}$ and $x_{1}-y_{1}.$ We note that when $H_{1}^{\prime}$ and/or
$\pm\tfrac{\partial}{\partial v_{1}}+i\tfrac{\partial}{\partial x_{1}}$ acts
upon $\left|  a\right\rangle $, the result is again a state constant in
$u_{1}-v_{1}$ and $x_{1}-y_{1}$. It therefore suffices to compute the
correlator restricted to the subspace which is constant in $u_{1}-v_{1}$ and
$x_{1}-y_{1}$. In this space $H_{1}^{\prime}$ has the form%
\begin{align}
H_{1}^{\prime}  &  \rightarrow-\tfrac{\alpha+1}{8t}\left(  \tfrac{\partial
}{\partial u_{1}}+\tfrac{\partial}{\partial v_{1}}\right)  ^{2}-\tfrac
{\alpha+1}{8t}\left(  \tfrac{\partial}{\partial x_{1}}+\tfrac{\partial
}{\partial y_{1}}\right)  ^{2}\\
&  -\tfrac{1}{t}\left[  \left(  u_{1}+v_{1}\right)  \left(  \tfrac{\partial
}{\partial u_{1}}+\tfrac{\partial}{\partial v_{1}}\right)  +\left(
x_{1}+y_{1}\right)  \left(  \tfrac{\partial}{\partial x_{1}}+\tfrac{\partial
}{\partial y_{1}}\right)  \right]  .\nonumber
\end{align}
Comparing with (\ref{ze}), we see that this is analogous with the previous
case for $n=0$. We again find the result%
\begin{equation}
\left\langle 0\right|  F_{1}^{12}(t)F_{-1}^{12}(t^{\prime})\left|
0\right\rangle =\tfrac{1}{t}\delta(t-t^{\prime}).
\end{equation}

\section{Summary}

In this work we applied the methods of spherical field theory to free gauge
fields. We analyzed two dimensional gauge fields in general covariant gauge
and radial gauge. In the process we have discussed several new aspects which
resulted from the spin degrees of freedom as well as the masslessness of the
gauge field. As we have seen, polarization mixing complicates the structure of
the spherical field Hamiltonian. Nevertheless in radial gauge we were able to
decompose the spherical field Hamiltonian as a sum of harmonic oscillators. We
did the same for covariant gauge, but found that the $s$-wave part of the
Hamiltonian has continuous spectrum. In relation to these differences, we also
discussed issues regarding the $s$-wave boundary condition at $t=0.$ We then
used the spherical field evolution equations to calculate two-point
correlators for the gauge fields and field-strength tensors $F^{12}$. Our
presentation here is intended as a first introduction to the application of
spherical field methods to gauge theories. Free gauge fields in higher
dimensions can be treated by a straightforward generalization of these
methods. Interacting gauge systems, however, include many interesting
theoretical and computational issues not discussed here, and these are the
subject of active research.

\end{document}